\documentclass{optica-article}

\journal{opticajournal} 

\articletype{Research Article}
\usepackage{lineno}
\usepackage{amsmath}
\usepackage{graphicx}
\usepackage{bm}
\usepackage{multirow}
\usepackage{booktabs}
\usepackage{float}
\usepackage{csquotes}

\usepackage{abraces}
\usepackage{commath}
\usepackage{ulem}
\usepackage{braket}
\usepackage{cleveref}
\usepackage{array}
\usepackage{xcolor}
\usepackage{soul}
\usepackage{cancel}

\begin{document}

\title{Entanglement swapping in high dimensions with only linear optics}

\author{Baghdasar Baghdasaryan,\authormark{1} Kaushik Joarder,\authormark{1} and Fabian Steinlechner\authormark{1,2,*}}

\address{\authormark{1}Friedrich Schiller University Jena, Institute of Applied Physics, Abbe Center of Photonics, 07745 Jena, Germany\\
\authormark{2}Fraunhofer Institute for Applied Optics and Precision Engineering IOF, 07745 Jena, Germany}

\email{\authormark{*}baghdasar.baghdasaryan@uni-jena.de} 

\begin{abstract*} 
Entanglement swapping is a fundamental building block for realizing first-generation quantum repeaters, which are essential for building global quantum networks. Current quantum repeater systems still struggle to achieve practical communication rates. High-dimensional (HD) encoding can significantly improve repeater efficiency by boosting information capacity and enhancing noise tolerance and security. However, the experimental demonstration of this protocol so far has been limited only to two-dimensional systems due to the requirement of strong nonlinear interactions. Here, we theoretically show that a modular linear-optics setup can implement HD entanglement swapping based on ancillary photons. For a four-dimensional scenario, we present an experimental design that employs hyper-entanglement in the polarization and time-bin degrees of freedom. This setup resolves the most challenging part of the ancillary photons-based approach, namely the necessary preparation of the ancilla state and the analysis of the resulting swapped state.

\end{abstract*}

\section{Introduction}
Large-scale quantum communication networks aim to distribute quantum states and quantum correlations between distant nodes \cite{PhysRevLett.81.5932,Duan2001,PhysRevLett.101.080403,Liao2017,doi:10.1126/science.aan3211}, enabling tasks such as quantum key distribution (QKD) \cite{Cui2019,Wang2021}, quantum secure direct communication (QSDC) \cite{PhysRevA.68.042317, PhysRevA.71.044305,SHENG2022367,10440135,10.1109/MNET.108.2100375,PhysRevLett.118.220501,10.1063/5.0245163,Paparelle:25,YANG20251445}, quantum teleportation \cite{PhysRevLett.70.1895} and entanglement swapping \cite{PhysRevLett.71.4287}. These tasks use quantum states and entanglement in different ways: QKD establishes shared secret keys, QSDC aims to transmit secret messages directly through quantum states without first generating a shared key, quantum teleportation uses entanglement to transfer unknown quantum states over a distance, and entanglement swapping relays entanglement across network nodes.

Since the first demonstration of quantum teleportation \cite{Bouwmeester1997} and entanglement swapping \cite{PhysRevLett.80.3891}, significant progress has been made to enhance these protocols \cite{Ghosh_2002, Jin2015, Hassanpour2016, Hu2023,doi:10.1126/sciadv.adf4080, Alipour2024, D’Aurelio2025}, including the generalization of the protocols to multiple particles, a necessary step for scaling quantum networks \cite{PhysRevA.57.822,PhysRevLett.103.020501,PhysRevLett.117.240503}. Most of these protocols have been realized using polarization- or path-entangled qubits, i.e., two-level quantum systems. However, it is well established that high-dimensional (HD) quantum states, qudits, offer several advantages over their two-dimensional counterparts. High-dimensional entangled states are bipartite states in which each photon is encoded in a Hilbert space of dimension $D>2$, rather than in a qubit subspace. A maximally entangled qudit state can be written as $\ket{\Phi_D}=\frac{1}{\sqrt D}\sum_{k=1}^{D}\ket{k,k}$, where the basis states $\ket{k}$ may correspond, for example, to time-bin, path, frequency, or orbital-angular-momentum modes. Compared with two-dimensional entanglement, such states possess higher information capacity \cite{Barreiro2008,PhysRevLett.108.143603,doi:10.1126/sciadv.aat9304,Nemirovsky-Levy:24,Nemirovsky-Levy2026}, greater robustness against eavesdropping, stronger violations of local realism \cite{advencies}, and enhanced potential for quantum computation \cite{Erhard2018}.

Photonic quantum technologies have made great progress, both as a scalable quantum information platform \cite{doi:10.1126/science.abe8770,Madsen2022} as well as the network-enabling ability to interconnect different processing nodes \cite{doi:10.1126/science.aam9288}. A key challenge in implementing HD variants of entanglement swapping \cite{Bouda_2001,Zhou_2009,zangi2025} or quantum teleportation based on photonic technologies lies in two main areas: the preparation and verification of HD entangled states \cite{PhysRevA.98.060301,PhysRevA.98.062316,PhysRevA.106.063711,PhysRevA.109.023534,9qrm-chgg,PhysRevA.110.033718}, and the realization of the Bell state measurement (BSM) \cite{PhysRevA.99.052301}. This latter difficulty arises because linear optics BSM typically relies on Hong-Ou-Mandel (HOM) interference on a two-port beam splitter (BS). Extending this approach to higher dimensions often necessitates the use of ancilla photons or nonlinear media \cite{PhysRevA.82.032318}. The first attempts were to teleport two-qubit composite systems \cite{Zhang2006}, teleport multiple degrees of freedom (DOFs) of a single photon \cite{Wang2015}, perform simultaneous entanglement swapping of different qubit systems in orbital angular momentum DOF \cite{Zhang2017} or frequency DOF  \cite{PhysRevLett.128.063602}.

Initial HD teleportation protocols were implemented in path DOF using ancilla photons to perform the HD BSM \cite{PhysRevLett.125.230501,PhysRevLett.123.070505}. Subsequent studies demonstrated that a BSM can be achieved through hyper-entanglement \cite{Zeng:22}, by using the Greenberger-Horne-Zeilinger state \cite{Lv2024}, by exploiting multiple squeezer devices \cite{PhysRevResearch.7.023038}, utilizing sum frequency generation \cite{qiu2021,Sephton2023}, by using quantum autoencoders to reconstruct qudit states \cite{doi:10.1126/sciadv.abn9783} or by using linear optics and permutation-entangled states as a resource \cite{PhysRevA.100.032330}. Recent work \cite{Bharos2025efficienthigh} discussed the utilization of quantum Fourier transform (QFT) with linear optics for implementing the HD BSM, tailored to time-bin encoding in even dimensions. However, the general implementation of $D$-dimensional QFT scales quadratically in the number of required BS operations \cite{Clements:16}. If the interference visibility is nonideal, the impact of these imperfections can accumulate with interferometric depth.

Despite significant progress, the experimental demonstration of HD entanglement swapping remains elusive. This is partly due to the lack of BSM approaches that can be tailored to
the specific challenges of time-bin, spatial mode, or frequency encoding. To address this, we introduce a BSM setup for qudit states based on linear optical quantum computing and ancillary photons in both odd and even dimensions. To keep the approach general, the theoretical scheme is formulated in terms of generic internal modes, without fixing a specific physical encoding. Compared with Ref. \cite{Bharos2025efficienthigh}, our scheme requires fewer BS operations, which offers a decisive advantage when HOM visibility is limited. In addition to the targeted high-dimensional swapped states, the scheme also produces two-dimensional entangled states, which may be useful in settings where both lower- and higher-dimensional entangled resources are relevant. Finally, we discuss a concrete experimental setup for realizing entanglement swapping with hyper-entangled states in time-bin \cite{Halder2007,PhysRevA.95.032306,Sun:17,Samara_2021} and polarization DOFs \cite{PhysRevA.77.022312,Tsujimoto2018,PhysRevLett.123.160502}.  This setup directly addresses the primary limitation of approaches utilizing ancillary photons, namely the complex preparation of the ancilla state.

This manuscript is organized into three main Sections. Section \ref{sec2A} outlines the general implementation of a three- to six-dimensional BSM. Following this, Section \ref{sec2B} proposes a specific implementation of four-dimensional BSM, leveraging hyper-entanglement across time-bin and polarization DOFs. 
Finally, Section \ref{conc} provides the conclusion.

\section{Scheme and analysis for HD entanglement swapping}\label{sec2A}
\subsection{Four-dimensional entanglement swapping}\label{sec2.1A}
\begin{figure}[htbp]
\centering\includegraphics[width=7cm]{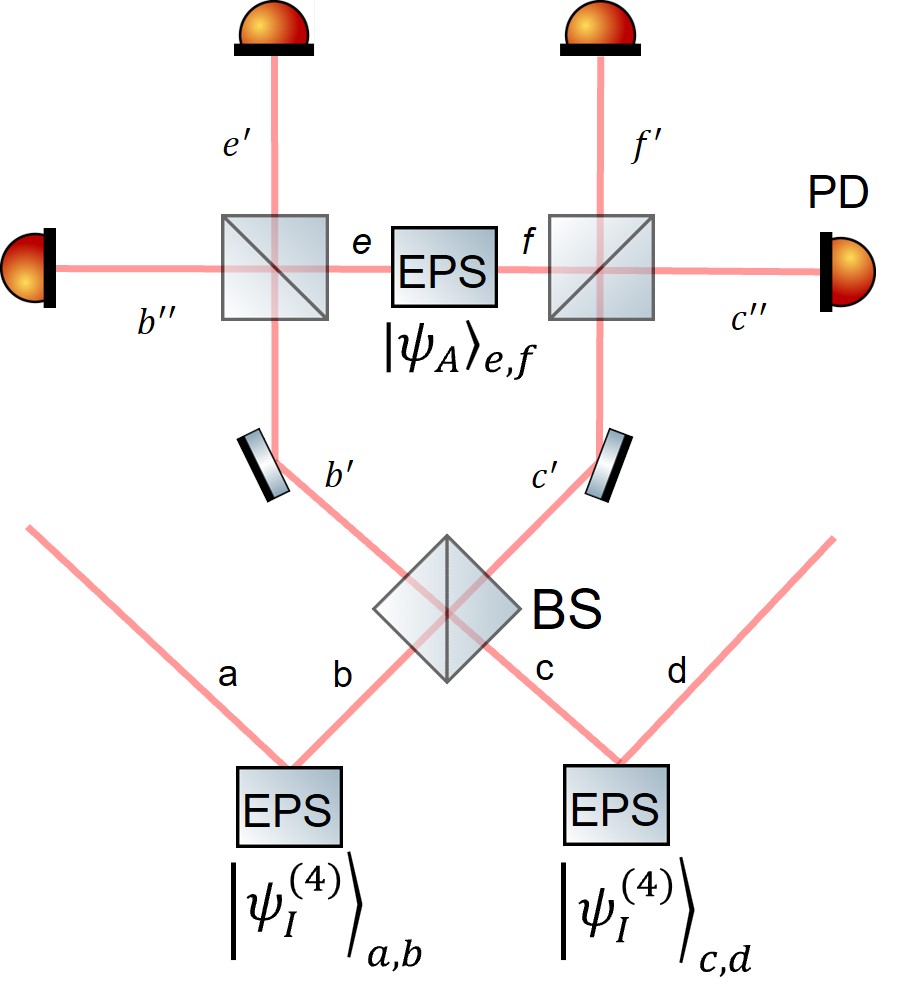}
\caption{Schematic of three- and four-dimensional entanglement swapping with three 50:50 BS operations (\textbf{BS}). In addition to two initially maximally entangled states, the setup also contains a two-dimensional ancilla state for performing the BSM. \textbf{EPS}: entangled photon-pair source; \textbf{PD}: photon detector.}\label{fig1}
\end{figure}

The schematic for four-dimensional entanglement swapping is shown in Fig. \ref{fig1}. The initial states can be any of $16$ orthogonal entangled Bell-like states in four dimensions \cite{Wang:17}. The derivation is shown only for
\begin{equation}
    \ket{\psi_I^{(4)}}_{m,n}=\frac{1}{2}\Bigl(\ket{1,1}_{m,n}+\ket{2,2}_{m,n}+\ket{3,3}_{m,n}+\ket{4,4}_{m,n}\Bigr),\label{state4}
\end{equation}
as it is analogous to all other four-dimensional Bell states. Here, $I$ refers to the initial state, the subscripts $m$ and $n$ denote the spatial modes, and the internal mode numbers $1-4$ correspond to the orthogonal states in any DOF in which the entanglement is encoded. The initial state consists of the first entangled state in spatial modes $a$ and $b$, denoted as $ \ket{\psi_I^{(4)}}_{a,b}$, and the second entangled state in spatial modes $c$ and $d$, $ \ket{\psi_I^{(4)}}_{c,d}$. A complete HD BSM cannot be implemented with linear optics only \cite{PhysRevA.65.030301}. Therefore, in addition to the initial states, we consider a maximally entangled ancillary state, which is prepared in spatial modes $e$ and $f$, $\ket{\psi_A}_{e,f}$. Therefore, the composite state of all three subsystems reads as $ \ket{\psi_I^{(4)}}_{a,b} \ket{\psi_I^{(4)}}_{c,d}\ket{\psi_A}_{e,f}$.

In the first step, we apply a $50:50$ BS operation on photons in paths $b$ and $c$. Let us write explicitly the BS transformation of the term  $\ket{1,1}_{a,b}\ket{2,2}_{c,d}$. Since the BS acts only on modes $b$ and $c$, the modes $a$ and $d$ remain unaffected. The transformation of the modes $\ket{1}_{b}\ket{2}_{c}=b_{1}^{\dagger}c_{2}^{\dagger}\ket{vac}$ follows the BS transformation rule \cite{Gerry_Knight_2004}:
\begin{equation}
   \begin{pmatrix} {b_k}^{\dagger} \\ {c_k}^{\dagger}\end{pmatrix}=  \frac{1}{\sqrt{2}} \begin{pmatrix} 1 & i \\ i & 1 \end{pmatrix}\begin{pmatrix} {b_k^{\prime}}^{\dagger} \\  {c_k^{\prime}}^{\dagger}\end{pmatrix}.\label{matrix}
\end{equation}
If we substitute $b_{1}^{\dagger}c_{2}^{\dagger}\ket{vac}=\frac{1}{\sqrt2}\bigl(b_1'{}^\dagger+ic_1'{}^\dagger\bigr)\,\frac{1}{\sqrt2}\bigl(ib_2'{}^\dagger+c_2'{}^\dagger\bigr)$, the whole transformation results in:
\begin{align*} \ket{1,1}_{a,b}&\ket{2,2}_{c,d}\overset{\text{BS}}{\rightarrow} \frac{1}{2}\ket{1}_a\ket{2}_d\,\Bigl(\ket{1}_{b^{\prime}}\ket{2}_{c^{\prime}}-\ket{2}_{b^{\prime}}\ket{1}_{c^{\prime}} \\&+\,i\ket{1}_{b^{\prime}}\ket{2}_{b^{\prime}}\ket{vac}_{c^{\prime}} +i\ket{vac}_{b^{\prime}}\ket{1}_{c^{\prime}}\ket{2}_{c^{\prime}}\Bigl).
\end{align*}
The notation $\ket{2}_{c^{\prime}}$ means a single photon in the spatial mode $c^{\prime}$ with internal DOF of $2$ and should not be confused with Fock state notation. We initially employed the explicit notation $\ket{1}_{b^{\prime}}\ket{2}_{b^{\prime}}$ to emphasize the presence of two photons within the single spatial mode $b^{\prime}$, distinguished by their internal mode indices $1$ and $2$. For the remainder of this work, we adopt the compact notation $\ket{k}_m\ket{j}_{n} \equiv \ket{k,j}_{m,n}$.

As in the two-dimensional entanglement swapping, we post-select only those events where all four detectors click. In other words, we focus only on the terms where the photons exit the BS through different spatial modes. Therefore, from the transformation of $\ket{1,1}_{a,b}\ket{2,2}_{c,d}$, we only retain the following term:
\begin{equation*}   \ket{1,2}_{a,d}\,\Bigl(\ket{1,2}_{b^{\prime},c^{\prime}}-\ket{2,1}_{b^{\prime},c^{\prime}} \Bigl) = \ket{1,2}_{a,d}\, \ket{\psi_{1,2}^-}_{b^{\prime},c^{\prime}}.
\end{equation*}
Here, we have used the notation for the Bell state:
\begin{equation}
    \ket{\psi_{k,j}^-}_{m,n}=
    \frac{1}{\sqrt{2}}\Bigr(\ket{k,j}_{m,n}-\ket{j,k}_{m,n}\Bigr).\label{bellS}
\end{equation}
Similarly, the transformation of $\ket{2,2}_{a,b}\ket{1,1}_{c,d}$ yields the term $-\ket{2,1}_{a,d}\, \ket{\psi_{1,2}^-}_{b^{\prime},c^{\prime}}$. Combining these relevant modes ultimately results in $ \ket{\psi_{1,2}^-}_{a,d}\,\ket{\psi_{1,2}^-}_{b^{\prime},c^{\prime}}$. Since the beam splitter transforms only the spatial creation operators and does not mix the internal-modes, only terms of the form $\ket{i,i}_{a,b}\ket{j,j}_{c,d}$ with $i\neq j$ contribute to the post-selection with one photon in each of the output modes $b'$ and $c'$. In contrast, terms with identical modes in both input pairs, such as $\ket{1,1}_{a,b}\ket{1,1}_{c,d}$, correspond to two indistinguishable photons interfering on the first BS and lead only to terms where photons exit the same output.

Applying the same derivation steps for other mode combinations and post-selecting only the terms with distinct spatial mode outputs, the transformation of the two initial states $ \ket{\psi_I^{(4)}}_{a,b} \ket{\psi_I^{(4)}}_{c,d}$ after the first BS can be expressed in an unnormalized form as:

\begin{align}
 &\ket{\psi_{1,2}^-}_{a,d}\,\ket{\psi_{1,2}^-}_{b^{\prime},c^{\prime}}+
    \ket{\psi_{1,3}^-}_{a,d}\,\ket{\psi_{1,3}^-}_{b^{\prime},c^{\prime}}+
    \ket{\psi_{1,4}^-}_{a,d}\,\ket{\psi_{1,4}^-}_{b^{\prime},c^{\prime}}+
    \ket{\psi_{2,3}^-}_{a,d}\,\ket{\psi_{2,3}^-}_{b^{\prime},c^{\prime}}\nonumber\\&+
    \ket{\psi_{2,4}^-}_{a,d}\,\ket{\psi_{2,4}^-}_{b^{\prime},c^{\prime}}+
    \ket{\psi_{3,4}^-}_{a,d}\,\ket{\psi_{3,4}^-}_{b^{\prime},c^{\prime}}\label{first}.
\end{align}
After the first BS, the state of four photons is a coherent superposition of products of two-dimensional Bell states. To engineer a genuine four-dimensional entangled state in modes $a$ and $d$, we must coherently combine specific pairs of these singlets. For example, we might aim to combine the terms $\ket{\psi_{1,2}^-}_{a,d}$ and $\ket{\psi_{3,4}^-}_{a,d}$, which appear in the first and last terms of Eq. \eqref{first}, to have the four-dimensional maximally entangled state $ 1/\sqrt{2}\,\Bigr(\ket{\psi_{1,2}^-}_{a,d}+\ket{\psi_{3,4}^-}_{a,d}\Bigl)$. To achieve this coherent combination, the relevant terms must share a common product component, allowing us to factor out the desired superposition, such as $\Bigr(\ket{\psi_{1,2}^-}_{a,d}+\ket{\psi_{3,4}^-}_{a,d}\Bigr) \times (\text{common product})$. Given that the first term is accompanied by $\ket{\psi_{1,2}^-}_{b^{\prime},c^{\prime}}$ and the second with $\ket{\psi_{3,4}^-}_{b^{\prime},c^{\prime}}$, the $\ket{\psi_{1,2}^-}_{b^{\prime},c^{\prime}}$ component needs to interfere with ancillary mode $\ket{3,4}_{e,f}$, while the term $\ket{\psi_{3,4}^-}_{b^{\prime},c^{\prime}}$ with ancillary mode $\ket{1,2}_{e,f}$. This interference can be accomplished by preparing the ancillary state in a superposition of these modes:
\begin{equation}
\label{ancilla_4d}
\ket{\psi_A}_{e,f}=\frac{1}{\sqrt{2}}\Bigr(\ket{1,2}_{e,f}+e^{i\varphi}\ket{3,4}_{e,f}\Bigr),
\end{equation}
where we introduced a variable phase $\varphi$ for the ancilla state. We analyze the interference of the remaining state \eqref{first} with the ancilla state \eqref{ancilla_4d} as shown in Fig. \ref{fig1}. Because the beam splitter acts exclusively on the spatial creation operators, it leaves the internal degrees of freedom unmixed. Therefore, the heralding projection onto
$\ket{\psi^-_{1,3}}_{b'',e'}\ket{\psi^-_{2,4}}_{c'',f'}$
can only arise from contributions carrying the internal-mode set $\{1,2,3,4\}$. Among the six terms in Eq.~\eqref{first}, only the first and the last terms satisfy this condition. To demonstrate this, the two BS transformations are applied to these two terms in Eq. \eqref{first} (for a more detailed derivation see Appendix \ref{appemdix1}):
\begin{align*}
&\ket{\psi_{1,2}^-}_{a,d}\,\ket{\psi_{1,2}^-}_{b^{\prime},c^{\prime}}\ket{\psi_{A}}_{e,f}\overset{\text{BS}}{\rightarrow}  \frac{1}{4}\ket{\psi_{1,2}^-}_{a,d} \Bigr(-
\ket{\psi_{2,1}^-}_{b^{\prime\prime},e^{\prime}}\ket{\psi_{1,2}^-}_{c^{\prime\prime},f^{\prime}} +e^{i\varphi}\bm{\ket{\psi_{1,3}^-}_{b^{\prime\prime},e^{\prime}}\ket{\psi_{2,4}^-}_{c^{\prime\prime},f^{\prime}}}\\&\qquad\qquad\qquad\qquad\qquad\qquad\qquad\qquad\qquad -e^{i\varphi}\ket{\psi_{2,3}^-}_{b^{\prime\prime},e^{\prime}}\ket{\psi_{1,4}^-}_{c^{\prime\prime},f^{\prime}} \Bigr),\\&
\ket{\psi_{3,4}^-}_{a,d}\,\ket{\psi_{3,4}^-}_{b^{\prime},c^{\prime}}\ket{\psi_{A}}_{e,f}\overset{\text{BS}}{\rightarrow} \frac{1}{4} \ket{\psi_{3,4}^-}_{a,d} \Bigr(
\bm{\ket{\psi_{1,3}^-}_{b^{\prime\prime},e^{\prime}}\ket{\psi_{2,4}^-}_{c^{\prime\prime},f^{\prime}}} -\ket{\psi_{4,1}^-}_{b^{\prime\prime},e^{\prime}}\ket{\psi_{3,2}^-}_{c^{\prime\prime},f^{\prime}}\\&\qquad\qquad\qquad\qquad\qquad\qquad\qquad\qquad\qquad -e^{i\varphi}\ket{\psi_{4,3}^-}_{b^{\prime\prime},e^{\prime}}\ket{\psi_{3,4}^-}_{c^{\prime\prime},f^{\prime}}\Bigr).
\end{align*}
Among all possible Bell state combinations, only the term $\ket{\psi_{1,3}^-}_{b^{\prime\prime},e^{\prime}}\ket{\psi_{2,4}^-}_{c^{\prime\prime},f^{\prime}}$ appears twice in the brackets. This is a direct consequence of our ancillary state choice. For example, selecting the ancilla state $\frac{1}{\sqrt{2}}\Bigr(\ket{1,4}_{e,f}+\ket{2,3}_{e,f}\Bigr)$ would instead lead to the term  $\ket{\psi_{1,2}^-}_{b^{\prime\prime},e^{\prime}}\ket{\psi_{4,3}^-}_{c^{\prime\prime},f^{\prime}}$ appearing twice. Through post-selection of this specific term, the unnormalized composite state is then:
\begin{equation*}
\Bigr(e^{i\varphi}\ket{\psi_{1,2}^-}_{a,d}+\ket{\psi_{3,4}^-}_{a,d}\Bigr)\,\ket{\psi_{1,3}^-}_{b^{\prime\prime},e^{\prime}}\ket{\psi_{2,4}^-}_{c^{\prime\prime},f^{\prime}}.
\end{equation*}
Therefore, when photons in spatial modes $b^{\prime\prime}$ and $e^{\prime}$ are projected onto the Bell state $\ket{\psi_{1,3}^-}_{b^{\prime\prime},e^{\prime}}$ and photons in spatial modes $c^{\prime\prime}$ and $f^{\prime}$ are projected onto $\ket{\psi_{2,4}^-}_{c^{\prime\prime},f^{\prime}}$, the state of remaining photons in spatial modes $a$ and $d$ becomes the four-dimensional maximally entangled state:
\begin{equation*}
\frac{1}{2}\Bigr(e^{i\varphi}\ket{1,2}_{a,d}-e^{i\varphi}\ket{2,1}_{a,d}+\ket{3,4}_{a,d}-\ket{4,3}_{a,d}\Bigr)
\end{equation*}
up to a global phase. The projection into these Bell states is achieved by measuring four-photon coincidences with four possible mode combinations $(\{3,1,4,2\},\{3,1,2,4\},\{1,3,4,2\}$ and $\{1,3,2,4\})$ in detectors at outputs $\{b^{\prime\prime},e^{\prime},f^{\prime},c^{\prime\prime}\}$. Any other four-photon coincidence event projects the state of photons in modes $a$ and $d$ into a two-dimensional Bell state. Note that the relative phase between modes $\Bigr(\ket{1,2}_{a,d}-\ket{2,1}_{a,d}\Bigr)$ and $\Bigr(\ket{3,4}_{a,d}-\ket{4,3}_{a,d}\Bigr)$ can be controlled by adjusting the phase $\varphi$ in the ancilla state.

\begin{figure}[htbp]
\centering\includegraphics[width=7cm]{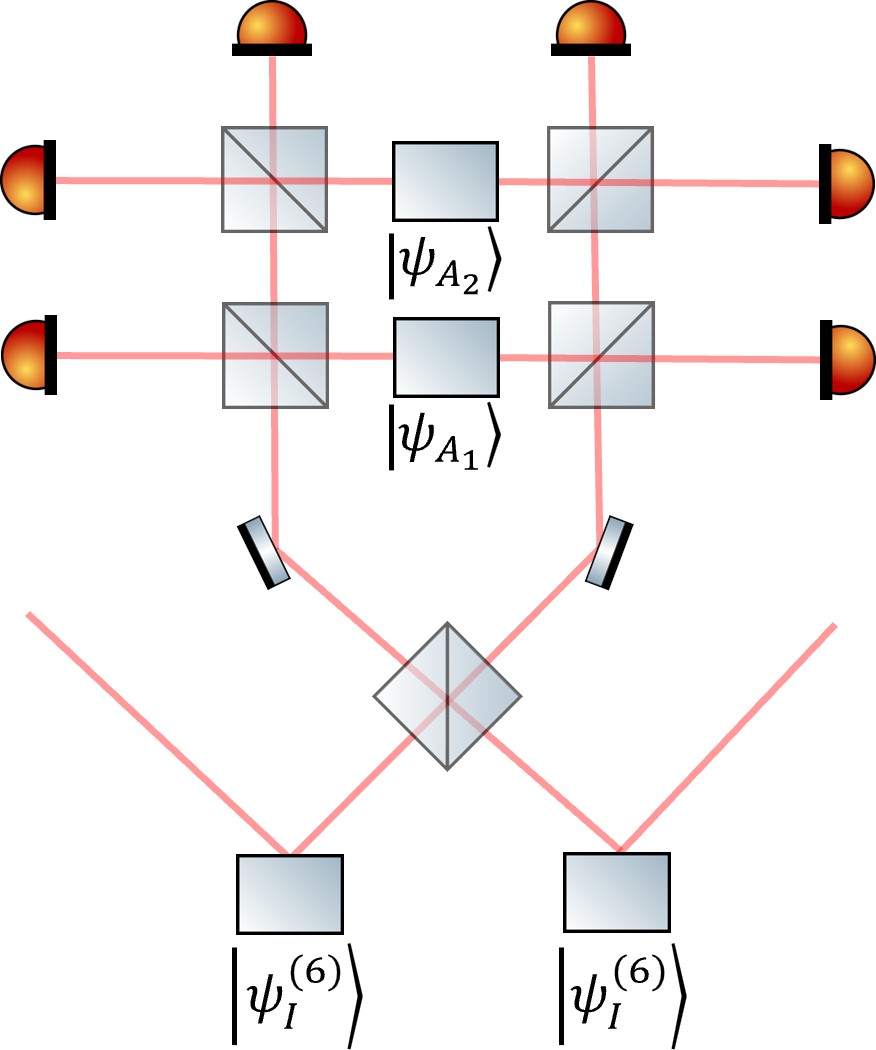}
\caption{Schematic of five- and six-dimensional entanglement swapping with two two-dimensional ancilla Bell states.}\label{fig2}
\end{figure}

The interferometric scheme illustrated in Fig. \ref{fig1} can be implemented on programmable photonic processors. To determine the unitary matrix, we express the input creation operators in terms of the output modes using the unitary transfer matrix of the network. By applying the beam splitter transformation from Eq. \eqref{matrix} to the specific topology in Fig. \ref{fig1}, we derive the following relation between the input and output modes:
\begin{equation*}
\renewcommand{\arraystretch}{1.2} 
 \begin{pmatrix}  {a_k}^{\dagger}\\ {b_k}^{\dagger}\\ {c_k}^{\dagger}\\ {d_k}^{\dagger}\\{e_k}^{\dagger} \\ {f_k}^{\dagger}\end{pmatrix}  =  \begin{pmatrix} 1&0&0&0&0&0 \\ 0 & \frac{1}{2}& \frac{i}{2} & 0 & \frac{i}{2} & -\frac{1}{2}\\ 0 & \frac{i}{2}& \frac{1}{2} & 0 & -\frac{1}{2} & \frac{i}{2} \\
   0 & 0& 0 & 1 & 0 & 0 \\
   0 & \frac{i}{\sqrt{2}}& 0 & 0 &  \frac{1}{\sqrt{2}} & 0\\
   0 & 0& \frac{i}{\sqrt{2}} & 0 &  0 & \frac{1}{\sqrt{2}} \end{pmatrix}\begin{pmatrix}  {a_k}^{\dagger}\\ {b_k^{\prime\prime}}^{\dagger}\\ {c_k^{\prime\prime}}^{\dagger}\\ {d_k}^{\dagger}\\{e_k^{\prime}}^{\dagger} \\ {f_k^{\prime}}^{\dagger}\end{pmatrix}.
\end{equation*}
\subsection{Three-dimensional entanglement swapping}
The same architecture in Fig. \ref{fig1} can also realize three-dimensional entanglement swapping, which reduces the number of occupied modes and thereby increases the overall success probability. The ancilla state remains unchanged, but the initial three-dimensional states must be changed to:
\begin{equation}
  \ket{\psi_I^{(3)}}_{m,n}=\frac{1}{\sqrt{3}}\Bigr(\ket{1,2}_{m,n}+\ket{2,1}_{m,n}+\ket{3,4}_{m,n}\Bigr),\label{odd1}
\end{equation}
or
\begin{equation}
    \ket{\psi_I^{(3')}}_{m,n}=\frac{1}{\sqrt{3}}\Bigr(\ket{1,1}_{m,n}+\ket{2,2}_{m,n}+\ket{3,4}_{m,n}\Bigr),\label{odd2}
\end{equation}
where the relative phase between modes can be chosen arbitrarily. These states are three-dimensional maximally entangled, but embedded in a four‑dimensional space.

Let us consider the input state in Eq.~\eqref{odd2}
together with the same ancillary state as in Eq.~\eqref{ancilla_4d}.
The full six-photon input state is therefore $
\ket{\psi_I^{(3')}}_{a,b}
\ket{\psi_I^{(3')}}_{c,d}
\ket{\psi_A}_{e,f}$.
The derivation for the three-dimensional entanglement swapping follows the same steps as the four-dimensional scenario. Applying the same BS transformations as in Sec. \ref{sec2.1A} and retaining only terms with one photon in each of the six output spatial modes gives 68 six-photon detection terms. Among these, four herald the three-dimensional state in modes $a$ and $d$. Similar to Sec. \ref{sec2.1A}, the successful projection is achieved by measuring four-photon coincidences with four possible mode combinations $(\{3,1,4,2\},\{3,1,2,4\},\{1,3,4,2\}$ and $\{1,3,2,4\})$ in detectors at outputs $\{b^{\prime\prime},e^{\prime},f^{\prime},c^{\prime\prime}\}$, which lead to the normalized three-dimensional maximally entangled state:
\begin{equation}
    \ket{\psi}_{a,d}=\frac{1}{\sqrt{3}}\Bigr(\ket{1,2}_{a,d}-\ket{2,1}_{a,d}-\ket{3,4}_{a,d}\Bigr),\label{anti}
\end{equation}
up to a global phase and for ancilla phase $\varphi=0$. Even though the initial states are defined in a four-dimensional subspace, the conditional efficiency for the three-dimensional entanglement swapping is higher than that for the four-dimensional scenario. 
\subsection{Five- and six-dimensional entanglement swapping}
The scheme in Fig. \ref{fig1} can be extended to higher dimensions. For instance, Fig. \ref{fig2} illustrates the setup for five- or six-dimensional entanglement swapping, which now requires two entangled ancilla states. The initial states for the six-dimensional scenario are again the generalized Bell states, for instance:
\begin{align*}
    \ket{\psi_I^{(6)}}_{m,n}=\frac{1}{\sqrt{6}}(&\ket{1,1}_{m,n}+\ket{2,2}_{m,n}+\ket{3,3}_{m,n}\\&+\ket{4,4}_{m,n}+\ket{5,5}_{m,n}+\ket{6,6}_{m,n}).
\end{align*}
The ancilla states are two-dimensional maximally entangled states $\ket{\psi_{A_1}}_{m,n}=1/\sqrt{2}(\ket{1,2}_{m,n}+\ket{3,4}_{m,n})$ and $\ket{\psi_{A_2}}_{m,n}=1/\sqrt{2}(\ket{5,6}_{m,n}+\ket{3,4}_{m,n})$. To achieve six-dimensional entanglement swapping, all interfered photons must be detected in distinct modes, as in the previous example. As expected, the scheme in Fig. \ref{fig2} is less efficient compared to the four-dimensional case. The total number of possible detection events is 106632, out of which only four lead to the desired entangled state, making its implementation in the near future very challenging. The generalized scheme for higher $D$-dimensions is presented in Appendix \ref{appendixB}, as its implementation would result in even lower efficiency.

\subsection{Discussion on success probability}
\label{sec24}%
To estimate the total efficiency of the four-dimensional entanglement swapping, the normalization factor of the final heralded state can be calculated. Each of the two initial entangled four-dimensional states carries a factor of $1/2$. The scheme uses one ancillary Bell pair normalized by $1/\sqrt{2}$. The number of BS operations is $3$, where each operation acts on two photons and introduces a factor of $1/2$. The overall normalization factor reads as
\begin{equation*}
     \Biggl(\frac{1}{\sqrt{2}\,2^{4}}\Biggl)\frac{1}{2}\Bigr(\ket{1,2}_{a,d}-\ket{2,1}_{a,d}+\ket{3,4}_{a,d}-\ket{4,3}_{a,d}\Bigr).
\end{equation*}
The last term $1/2$ is retained as the normalization factor of the heralded four-dimensional state. The probability of a single, specific heralded event is the square of the amplitude within the brackets, $1/2^9$. As four possible measurement outcomes successfully herald a four-dimensional entangled state in modes $a$ and $d$, the total success probability is $4/2^9$. An analogous calculation for the three-dimensional scenario delivers $1/96$ that is larger than the corresponding four-dimensional success probability $P_{\rm succ}^{(4)}=4/2^9=1/128$.

Inherently, a significant fraction of photons is lost since the scheme relies on ancilla photons and only considers events where photons exit into different spatial modes. Nevertheless, an attractive aspect of this scheme is that the remaining four-photon events still herald useful two-dimensional Bell states. For instance, the detection event with mode numbers $\{2,1,1,2\}$ leads to the Bell state $\ket{\psi_{1,2}^-}_{a,d}$, which follows from the first term in the multiline expression after Eq. \eqref{ancilla_4d}. There are in total $48$ four-photon events leading to two-dimensional swapping with success probability $48/2^{10}$. This feature could be useful in scenarios where both two-dimensional and four-dimensional states are required for different protocols.

Our setup and the scheme of Ref. \cite{Bharos2025efficienthigh} are conceptually similar and can, in principle, be implemented in different photonic DOFs. In the specific case of time-bin encoding, Ref. \cite{Bharos2025efficienthigh} outperforms our scheme in terms of success probability. However, this advantage is obtained in a detection scenario in which successful events must also involve multiple clicks on the same detector. In a time-bin implementation, this requires the separation between time bins to exceed the detector recovery (dead) time. For encodings other than time-bin, events in which multiple photons occupy the same spatial output mode generally require not only photon-number-resolving detection, but also a readout that resolves the internal photonic DOF. Realizing such measurements typically requires additional encoding-specific conversion or analysis stages, which makes the implementation substantially more demanding. For this reason, we adopt here a more experimentally accessible regime and restrict attention to heralding events in which all photons exit in distinct spatial modes.

Under this condition, the four-dimensional setup in Ref.~\cite{Bharos2025efficienthigh} yields a success probability of $24/2^{11}$. In our setup, the success probability for four-dimensional swapping is $1/128$. Hence, their approach offers a slightly higher probability for the strictly four-dimensional case. However, within the same experimentally motivated comparison, namely the distinct-spatial-output condition, the present scheme also produces two-dimensional entangled states, whereas in our analysis of Ref.~\cite{Bharos2025efficienthigh} we do not obtain corresponding two-dimensional swapping events under this restriction. Therefore, if both two- and four-dimensional entangled outputs are considered useful resources, 
the total per-attempt success probability of the present scheme is
$P_{succ}^{(2)}+P_{succ}^{(4)}=48/2^{10}+1/128=7/128$. 

This value represents an ideal theoretical limit. In a realistic experiment, the observed probability is reduced by transmission and detection losses. As a simple estimate, we model photon loss as an erasure channel and combine all losses for each heralding photon into an efficiency $\eta$. Since a successful heralding event requires four detected photons, the observed heralding probability becomes $P_{\rm succ}^{\rm ideal}\,\eta^4$. For example, for $\eta=0.8$, the heralding probability is reduced to $7/128\times 0.8^4 \approx 2.24\times 10^{-2}$. Since the detection of the two swapped photons in modes $a$ and $d$ is also required, their corresponding transmission/detection efficiencies must be included separately. The photon loss mainly reduces the experimentally observed rate, while in the ideal erasure model, the conditional state fidelity is unchanged, assuming that dark counts, mode-dependent losses, and multipair emissions are neglected.

If PNR detection with a mode-resolved readout becomes practically possible, the success probability of the protocol can be significantly enhanced. With PNR detection, the number of events leading to four-dimensional entanglement swapping is $16$ and to two-dimensional entanglement swapping is $552$, corresponding to success probabilities of $16/2^9$ for four-dimensional and $552/2^{10}\approx 0.54$ for two-dimensional entanglement swapping.

With two-dimensional input states, passive elements, no ancilla or feed-forward, and threshold detection, a BSM can unambiguously distinguish at most two of the four Bell states, bounding the success probability at $\leq 50\%$. Notably, the probability $ 0.54$ for two-dimensional swapping surpasses the conventional $50\%$ limit. This enhancement arises because the initial states are high-dimensional, so that after the first BS, the protocol already allows two-dimensional swapping with a higher probability. This mechanism has already been discussed in Ref.~\cite{PhysRevLett.134.200801} in the context of HD-assisted two-dimensional entanglement generation.

Beyond success probability, the fidelity of the final state is a critical metric too. In realistic implementations, HOM-based schemes can suffer from fidelity degradation due to imperfect indistinguishability. Imperfect HOM is primarily determined by the indistinguishability of the interfering photons. If the interference at each stage is not ideal, the effect of these imperfections can accumulate in deeper interferometric networks. Although a smaller number of beam splitters does not by itself guarantee improved mode matching, it reduces the number of interference stages and can therefore be advantageous from an experimental perspective. A complete quantitative fidelity model would require a source-specific treatment of multi-photon interference, including spectral-temporal distinguishability, detector timing jitter, photon coherence time, losses, detection efficiencies, and accidental coincidences~\cite{Baghdasaryan2026asynchronous}.

For $D=4$ dimension, our setup requires only 3 beam splitters compared to 4 in Ref. \cite{Bharos2025efficienthigh}. This advantage becomes significantly more pronounced in higher dimensions. For $D=6$ dimension, our scheme requires only 5 beam splitters. In contrast, a 6-mode QFT using standard universal linear optical decompositions can be realized with $D(D-1)/2=15$ BS operations \cite{Clements:16}. Consequently, there is an inherent trade-off between achieving a high success probability in HD BSM and maintaining the fidelity of the final state. 

The previous ancilla-assisted HD teleportation Ref.~\cite{PhysRevLett.125.230501} and our schemes are closely related at the resource level, since both use high-dimensional photonic states together with ancillary entangled two-dimensional photon pairs. However, that work experimentally demonstrates qutrit quantum teleportation and does not provide a general success-probability formula for an arbitrary dimension. Therefore, a quantitative success probability comparison for arbitrary $D$ cannot be made on equal footing. The schemes also differ in the structure of the Bell-state-measurement circuit. Ref.~\cite{PhysRevLett.125.230501} maps the $D$-dimensional qudit modes onto an auxiliary two-level sorting DOF, realized as polarization, before the parity check with the two-dimensional ancilla. In contrast, our central interferometer consists only of 50:50 beam splitters acting on spatial modes.

\section{Proposal: experimental setup for four-dimensional entanglement swapping}\label{sec2B}
We consider a four-dimensional entangled state encoded in a hyper-entangled form, using both time-bin and polarization DOFs. The use of hyper-entanglement makes the preparation of ancillary states and the verification of the final swapped state experimentally practical. Moreover, the use of hyper-entangled states enables the manipulation and measurement of each DOF independently with well-established linear optical techniques.

\begin{figure}[htbp]
\centering\includegraphics[width=\textwidth]{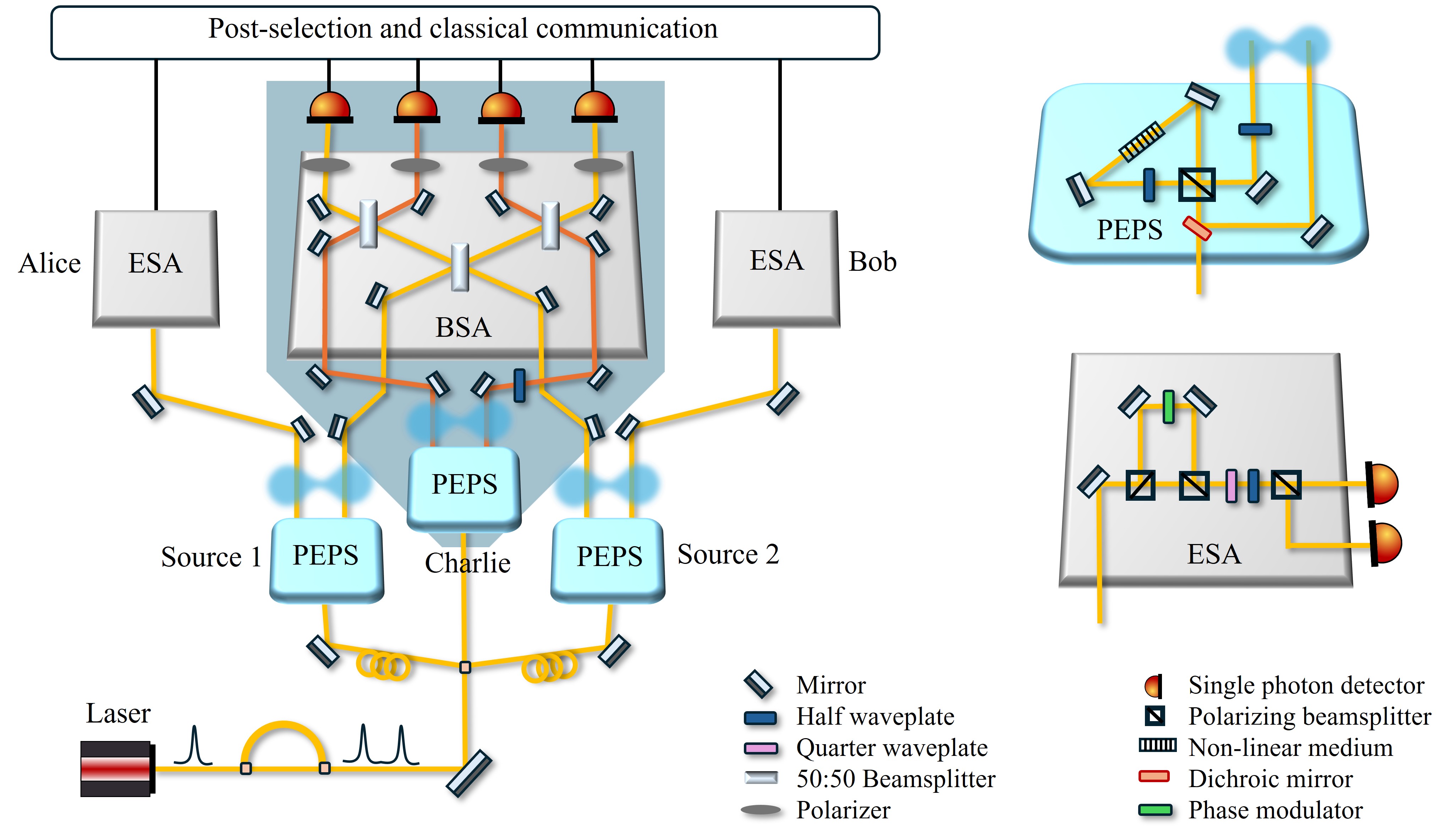}
\caption{Proposed experimental setup for four-dimensional entanglement swapping using hyper-entanglement in time-bin and polarization DOFs. \textbf{PEPS}: polarization-entangled photon-pair source; \textbf{BSA}: Bell state analyzer; \textbf{ESA}: entangled state analyzer.}\label{fig3}
\end{figure}

A four-dimensional hyper-entangled state in both sources 1 and 2 (Fig. \ref{fig3}) is prepared by using a polarization-entangled pair source (PEPS) pumped with two mutually coherent pulses: early and late pulses. The output photon pair becomes the hyper-entangled state:
\begin{equation}
     \ket{\psi_I^{(4)}}_{m,n}=\frac{1}{2}\Bigr(\ket{H(t_e),H(t_e)}_{m,n}+\ket{V(t_e),V(t_e)}_{m,n}+\ket{H(t_l),H(t_l)}_{m,n}+\ket{V(t_l),V(t_l)}_{m,n}\Bigr).\label{hyperstate}
\end{equation}
Here $t_{e(l)}$ represents the early (late) time-bin state of the photon, and $\ket{H(V)}$ shows the horizontal (vertical) polarization state.
The state in Eq. \eqref{hyperstate} is equivalent to the four-dimensional entangled state in Eq. \eqref{state4}, once we assign the values $\ket{1}=\ket{H(t_{e})}$, $\ket{2}=\ket{V(t_{e})}$, $\ket{3}=\ket{H(t_{l})}$, $\ket{4}=\ket{V(t_{l})}$.

To ensure optimal time synchronization and photon indistinguishability within the Bell state analyzer (BSA), a single master laser pumps the PEPSs for sources 1 and 2, as well as Charlie's ancilla state. In our visualization, mutually coherent pulse pairs are prepared using an asymmetric interferometer placed after the master laser. The pump beam is then equally split into three pulse pairs to pump the three PEPSs. Each PEPS can be readily constructed, for example, using three-wave mixing in a nonlinear medium \cite{10.1063/5.0023103}.
As a simple example, we consider a Sagnac-interferometer-based PEPS setup (as shown in Fig. \ref{fig3}) where a diagonally polarized ($+45^{\circ}$ or $-45^{\circ}$) pump beam with any temporal mode (early or late) splits into two orthogonally polarized counter-propagating beams after the polarizing beam splitter (PBS) and simultaneously pumps the non-linear crystal. Here we show a setup suitable for a type-II non-linear crystal, where a horizontally polarized pump photon creates a horizontally polarized signal and vertically polarized idler photon pair ($|H\rangle_{pump}\rightarrow|H\rangle_{sig}\,|V\rangle_{idl}$) via spontaneous parametric down conversion (SPDC) process. Both photon pair states created from counter-propagating pump beam recombine at the PBS, creating the Bell state, $\frac{1}{\sqrt{2}}(\ket{H,V}_{m,n}+\ket{V,H}_{m,n})$. The PBS also separates the signal from the idler photon, where an additional half-waveplate flips the polarization of one of the photons, creating the desired Bell state, $\frac{1}{\sqrt{2}}(\ket{H,H}_{m,n}+\ket{V,V}_{m,n})$.

We consider the ancillary state $\ket{\psi_A}_{m,n}=\frac{1}{2}\Bigr(\ket{1,2}_{m,n}+\ket{2,1}_{m,n}+\ket{3,4}_{m,n}+\ket{4,3}_{m,n}\Bigr)$, which is just a symmetric four-dimensional form of the state \eqref{ancilla_4d}, and does not change the derivation presented in Sec. \ref{sec2A}. This ancillary state is equivalent to the hyper-entangled state
\begin{equation}
    \ket{\psi_A}_{m,n}=\frac{1}{2}\Bigr(\ket{H(t_e),V(t_e)}_{m,n}+\ket{V(t_e),H(t_e)}_{m,n}+\ket{H(t_l),V(t_l)}_{m,n}+\ket{V(t_l),H(t_l)}_{m,n}\Bigr).\label{hyperstate2}
\end{equation}
Preparing the ancilla state becomes significantly easier in a hyper-entangled system. This is because any polarization-entangled two-photon qubit state can be transformed into any desired pure state within the two-qubit polarization subspace using only linear optical elements. In our setup, the ancilla state in Eq. \eqref{hyperstate2} can be prepared from state \eqref{hyperstate}, just by flipping the polarization state of the photons in spatial mode $n$ with a half-waveplate. 

The BSA setup is identical to Fig. \ref{fig1} for hyper-entanglement, with the addition of a polarizer in each of the four arms. These polarizers have a fixed projection angle, enabling post-selection alongside temporal mode selection. For example, to post-select the output state $\{3,1,4,2\}$ or equivalently $\{H(t_l),H(t_e),V(t_l),V(t_e) \}$ in a hyper-entangled form, we have to fix the polarizer angle of the first two polarizers in $H$ at $0^{\circ}$ and the last two polarizers in $V$ at $90^{\circ}$ direction. Also, we only consider those events when four photons are detected at time $\{t_l,t_e,t_l,t_e\}$ in the four detectors, respectively.

The final swapped state between Alice and Bob can be characterized using entangled state analyzers (ESA). Each ESA consists of a postselection-free Franson interferometer \cite{strekalov1996postselection}. Ideally, one may perform a full tomography setup as an ESA, which requires measurements in all five mutually unbiased bases (MUBs). However, as the dimensionality increases, the complexity of the setup and the total number of measurements also increase. Alternatively, one can measure a witness of entanglement \cite{PhysRevX.9.041042,bulla2023nonlocal}, where one can verify entanglement by checking correlations in the computational basis as well as in superpositions. Correlation checking in the computational basis is trivial, as one can measure the time of arrival and the polarization state of each photon pair at the ESA output. On the other hand, for the superposition measurement, additional steps should be considered. For example, to check the superposition between the states $\ket{V(t_e),H(t_e)}_{m,n}$ and $\ket{H(t_l),V(t_l)}_{m,n}$, one has to perform tomography in the polarization DOF after the Franson interferometer, between coincidence photon pairs detected at time $t_l$. On the other hand, to check the superposition between the states $\ket{H(t_e),V(t_e)}_{m,n}$ and $\ket{V(t_e),H(t_e)}_{m,n}$, one has to remove the asymmetric Mach-Zehnder interferometers inside ESAs and perform tomography in polarization DOF between coincidence photon pairs detected at time $t_e$.
\section{Conclusion}\label{conc}
In this work, we presented a scheme for implementing HD entanglement swapping in odd and even dimensions using only linear optical elements and ancillary photons. The work \cite{Bharos2025efficienthigh} proposed a related approach based on the QFT for HD Bell state analysis, tailored to time-bin encoding in even-dimensional systems. The schemes presented here require fewer BS operations, which can be a decisive advantage when the HOM visibility is limited. A trade-off exists between maximizing the success probability of HD BSM and preserving the fidelity of the final state. Within the experimentally motivated detection model adopted in Section \ref{sec24} — heralding only on events in which all photons exit in distinct spatial modes — our scheme produces an additional set of two-dimensional swapped outputs, so that the total per-attempt rate of usable entangled outputs is higher than that of Ref. \cite{Bharos2025efficienthigh}. However, both schemes suffer from an exponential suppression with increasing $D$ under the distinct-spatial-output restriction. Finally, we proposed a feasible experimental setup for realizing four-dimensional hyper-entanglement swapping using time-bin and polarization DOFs. This setup addresses the primary limitation of approaches utilizing ancillary photons: the preparation of the ancilla state and the subsequent analysis of the swapped state.
\appendix
\section{Appendix: Derivation of the BS transformation with ancilla state}\label{appemdix1}
We consider the transformation of the term
\begin{equation*}
\ket{\psi^-_{1,2}}_{a,d}\,\ket{\psi^-_{1,2}}_{b',c'}\,\ket{\psi_A}_{e,f}
\end{equation*}
under the two BSs acting on the pairs of spatial modes $(b',e)$ and $(c',f)$. Since these operations do not act on the outer modes $(a,d)$, the state $\ket{\psi^-_{1,2}}_{a,d}$ is unaffected and will be carried along as an overall factor. We use the definition in Eq. \eqref{bellS} for $\ket{\psi_{k,j}^-}_{m,n}$ and write the product on modes $(b',c',e,f)$ explicitly. For the present term, we have
\begin{equation*}
\ket{\psi^-_{1,2}}_{b',c'}\ket{\psi_A}_{e,f}
=
\frac{1}{\sqrt2}\Bigl(\ket{1,2}_{b',c'}-\ket{2,1}_{b',c'}\Bigr)\frac{1}{\sqrt{2}}\Bigr(\ket{1,2}_{e,f}+\ket{3,4}_{e,f}\Bigr).
\label{eq:expand_bc}
\end{equation*}
In the following, we illustrate the contribution arising from the component
$\ket{3}_{e}\ket{4}_{f}$ contained in $\ket{\psi_A}_{e,f}$. The remaining components are treated analogously and yield the other terms in the final expression. For a 50:50 BS acting on spatial modes $(b',e)$ we use the standard transformation from Eq. \eqref{matrix} on single-photon basis states:
\begin{equation*}
\ket{i}_{b'}\ket{k}_{e}\ \xrightarrow{\mathrm{BS}_{b'e}}\ 
\frac{1}{2}\Bigl(
\ket{i,k}_{b'',e'}-\ket{k,i}_{b'',e'}
\Bigr)
\;+\;(\text{bunched terms}),
\label{eq:BS_rule}
\end{equation*}
and analogously for the modes $(c',f)$.
We retain only the terms where the photons exit from different output ports. Hence, the BS transformation of the first product yields:
\begin{equation*}
\ket{1}_{b'}\ket{2}_{c'}\ket{3}_{e}\ket{4}_{f}
\ \xrightarrow{\mathrm{BS},\,\text{no bunching}}\
\frac{1}{2}\ket{\psi^-_{1,3}}_{b'',e'}\ket{\psi^-_{2,4}}_{c'',f'}
\label{eq:first_mapping}
\end{equation*}
and for the second term:
\begin{equation*}
\ket{2}_{b'}\ket{1}_{c'}\ket{3}_{e}\ket{4}_{f}
\ \xrightarrow{\mathrm{BS},\,\text{no bunching}}\
\frac{1}{2}\ket{\psi^-_{2,3}}_{b'',e'}\ket{\psi^-_{1,4}}_{c'',f'}.
\label{eq:second_mapping}
\end{equation*}

Including the factor $\ket{\psi^-_{1,2}}_{a,d}$ and the remaining terms treated analogously, we arrive at the full transformation quoted in the main text:
\begin{align*}
\ket{\psi^-_{1,2}}_{a,d}\,\ket{\psi^-_{1,2}}_{b',c'}\ket{\psi_A}_{e,f}
\ \xrightarrow{\mathrm{BS}}\
\frac{1}{4}\ket{\psi^-_{1,2}}_{a,d}\Bigl(&
-\ket{\psi^-_{2,1}}_{b'',e'}\ket{\psi^-_{1,2}}_{c'',f'}
+\ket{\psi^-_{1,3}}_{b'',e'}\ket{\psi^-_{2,4}}_{c'',f'}
\\&-\ket{\psi^-_{2,3}}_{b'',e'}\ket{\psi^-_{1,4}}_{c'',f'}
\Bigr).
\end{align*}
The transformation of all terms in Eq. \eqref{first} reads as
\begin{align*}
&\ket{\psi_{1,2}^-}_{a,d}\,\ket{\psi_{1,2}^-}_{b^{\prime},c^{\prime}}\ket{\psi_{A}}_{e,f}\overset{\text{BS}}{\rightarrow}  \frac{1}{4}\ket{\psi_{1,2}^-}_{a,d} \Bigr(-
\ket{\psi_{2,1}^-}_{b^{\prime\prime},e^{\prime}}\ket{\psi_{1,2}^-}_{c^{\prime\prime},f^{\prime}} +e^{i\varphi}\bm{\ket{\psi_{1,3}^-}_{b^{\prime\prime},e^{\prime}}\ket{\psi_{2,4}^-}_{c^{\prime\prime},f^{\prime}}}\\&\qquad\qquad\qquad\qquad\qquad\qquad\qquad\qquad\qquad -e^{i\varphi}\ket{\psi_{2,3}^-}_{b^{\prime\prime},e^{\prime}}\ket{\psi_{1,4}^-}_{c^{\prime\prime},f^{\prime}} \Bigr),\\&
\ket{\psi_{1,3}^-}_{a,d}\,\ket{\psi_{1,3}^-}_{b^{\prime},c^{\prime}}\ket{\psi_{A}}_{e,f}
\overset{\text{BS}}{\rightarrow} \frac{1}{4} \ket{\psi_{1,3}^-}_{a,d} \Bigr(e^{i\varphi}
\ket{\psi_{1,3}^-}_{b^{\prime\prime},e^{\prime}}\ket{\psi_{3,4}^-}_{c^{\prime\prime},f^{\prime}} -\ket{\psi_{3,1}^-}_{b^{\prime\prime},e^{\prime}}\ket{\psi_{1,2}^-}_{c^{\prime\prime},f^{\prime}} \Bigr)
,\\&
\ket{\psi_{1,4}^-}_{a,d}\,\ket{\psi_{1,4}^-}_{b^{\prime},c^{\prime}}\ket{\psi_{A}}_{e,f}
\overset{\text{BS}}{\rightarrow} \frac{1}{4} \ket{\psi_{1,4}^-}_{a,d}  \Bigr(-
\ket{\psi_{4,1}^-}_{b^{\prime\prime},e^{\prime}}\ket{\psi_{1,2}^-}_{c^{\prime\prime},f^{\prime}} -e^{i\varphi}\ket{\psi_{4,3}^-}_{b^{\prime\prime},e^{\prime}}\ket{\psi_{1,4}^-}_{c^{\prime\prime},f^{\prime}}\Bigr ),\\&
\ket{\psi_{2,3}^-}_{a,d}\,\ket{\psi_{2,3}^-}_{b^{\prime},c^{\prime}}\ket{\psi_{A}}_{e,f}
\overset{\text{BS}}{\rightarrow} \frac{1}{4} \ket{\psi_{2,3}^-}_{a,d} \Bigr(
\ket{\psi_{2,1}^-}_{b^{\prime\prime},e^{\prime}}\ket{\psi_{3,2}^-}_{c^{\prime\prime},f^{\prime}} +e^{i\varphi}\ket{\psi_{2,3}^-}_{b^{\prime\prime},e^{\prime}}\ket{\psi_{3,4}^-}_{c^{\prime\prime},f^{\prime}} \Bigr),\\&
\ket{\psi_{2,4}^-}_{a,d}\,\ket{\psi_{2,4}^-}_{b^{\prime},c^{\prime}}\ket{\psi_{A}}_{e,f}
\overset{\text{BS}}{\rightarrow} \frac{1}{4} \ket{\psi_{2,4}^-}_{a,d} \Bigr(
\ket{\psi_{2,1}^-}_{b^{\prime\prime},e^{\prime}}\ket{\psi_{4,2}^-}_{c^{\prime\prime},f^{\prime}} -e^{i\varphi}\ket{\psi_{4,3}^-}_{b^{\prime\prime},e^{\prime}}\ket{\psi_{2,4}^-}_{c^{\prime\prime},f^{\prime}}\Bigr ),\\&
\ket{\psi_{3,4}^-}_{a,d}\,\ket{\psi_{3,4}^-}_{b^{\prime},c^{\prime}}\ket{\psi_{A}}_{e,f}\overset{\text{BS}}{\rightarrow} \frac{1}{4} \ket{\psi_{3,4}^-}_{a,d} \Bigr(
\bm{\ket{\psi_{1,3}^-}_{b^{\prime\prime},e^{\prime}}\ket{\psi_{2,4}^-}_{c^{\prime\prime},f^{\prime}}} -\ket{\psi_{4,1}^-}_{b^{\prime\prime},e^{\prime}}\ket{\psi_{3,2}^-}_{c^{\prime\prime},f^{\prime}}\\&\qquad\qquad\qquad\qquad\qquad\qquad\qquad\qquad\qquad -e^{i\varphi}\ket{\psi_{4,3}^-}_{b^{\prime\prime},e^{\prime}}\ket{\psi_{3,4}^-}_{c^{\prime\prime},f^{\prime}}\Bigr).
\end{align*}
\section{D-DIMENSIONAL ENTANGLEMENT SWAPPING}\label{appendixB}
\begin{figure}[htbp]
\centering\includegraphics[width=7cm]{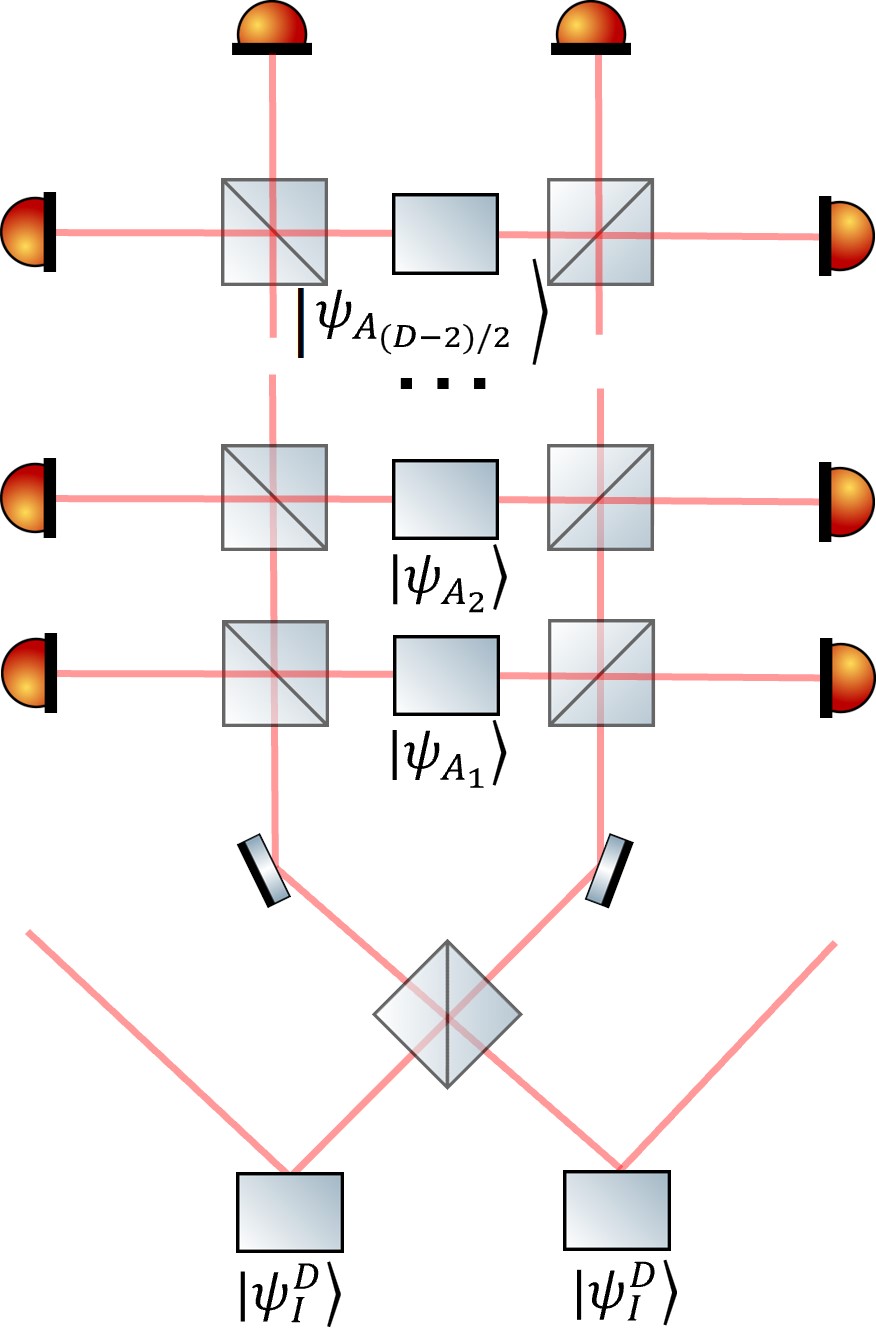}
\caption{Schematic of $D$- and $(D-1)$-dimensional
entanglement swapping with $(D-2)/2$ two-dimensional ancilla Bell states.}\label{fig4}
\end{figure}
The generalization of the schematic presented in Fig.~1 to $D$- and $(D-1)$-dimensional entanglement swapping is illustrated in Fig. \ref{fig4} where $D$ is an even number. The schematic in Fig.~\ref{fig4} is a modular extension of the four-dimensional setup. Entanglement swapping for both $D+1$ and $D+2$ dimensions can be performed if an additional layer of ancillary state is added to the existing $D$-dimensional schematic. The initial states in the even $D$-dimensional scenario can be any of the generalized Bell states in $D$ dimensions. The initial states for $(D-1)$ odd-dimensional scenario are generalized to
\begin{equation}
\ket{\psi_i^{D-1}}
=
\frac{1}{\sqrt{D-1}}
\left(
\sum_{j=1}^{D-2} \ket{j,j}
+
\ket{D-1,D}
\right),
\end{equation}
or to the $(D-1)$-dimensional generalization of the state \eqref{odd1}.

The number of ancilla states is the same for both $D$- and $(D-1)$-dimensional entanglement swapping. To perform the $D$-dimensional BSM, $(D-2)/2$ two-dimensional Bell states are required in the form of
\begin{equation}
\ket{\psi_{A_j}}
=
\frac{1}{\sqrt{2}}
\left(
\ket{2j-1,\,2j}
+
\ket{2j+1,\,2j+2}
\right),
\end{equation}
where $j$ denotes the $j$-th of $(D-2)/2$ ancilla states. The projection into the $D$-dimensional Bell state is achieved by measuring specific $D$-photon coincidences with $D$ distinct mode numbers. The conditional efficiency of the entanglement swapping decreases with increasing dimension due to increasing mode combinations. For instance, the total number of possible click patterns is $106632$ for $D=6$, out of which only four lead to the desired six-dimensional maximally entangled state. The remaining six click events lead either to four- or two-dimensional maximally entangled states.

In order to estimate the total efficiency of the $D$-dimensional entanglement swapping, the normalization factor of the final heralded state can be calculated as a function of $D$. Each of the two initial entangled $D$-dimensional states carries a factor of $1/\sqrt{D}$. The scheme uses $(D-2)/2$ ancillary Bell pairs, each normalized by $1/\sqrt{2}$. The number of beam splitter operations is $D-1$, where each operation acts on two photons and introduces a factor of $1/2$. The overall normalization factor reads as
\begin{equation}
\frac{1}{\sqrt{D}}
\frac{1}{\sqrt{D}}
\left( \frac{1}{\sqrt{2}} \right)^{\frac{D-2}{2}}
\left( \frac{1}{2} \right)^{D-1}
=
\left[
\frac{1}{\sqrt{D}}
\left( \frac{1}{2} \right)^{\frac{5D-6}{4}}
\right]
\frac{1}{\sqrt{D}}.
\end{equation}

The last term $1/\sqrt{D}$ is retained as the normalization factor of the heralded $D$-dimensional state. The probability of a single, specific heralded event is the square of the amplitude within the brackets. As four possible measurement outcomes successfully herald $D$-dimensional entanglement swapping, the total success probability is:
\begin{equation}
P_{\mathrm{succ}}(D)
=
\frac{4}{D}
\left( \frac{1}{2} \right)^{\frac{5D-6}{2}}.
\end{equation}
Given the exponential decrease in success probability with dimension $D$, initial experimental efforts should prioritize the implementation of three- and four-dimensional entanglement swapping as a proof-of-principle demonstration. Here the scaling refers to the qudit dimension $D$, i.e. the local Hilbert-space dimension of each encoded photon, which is carried by a single photon irrespective of whether its 
$D$ levels come from one or several degrees of freedom. However, one may gain additional advantages in a real-world experimental implementation while using multiple DOF instead of a single DOF. The overall physical-resource cost $D+2$ photons, $D-1$ beam splitters, $(D-2)/2$ ancilla pairs, therefore grows only linearly in $D$.

\begin{backmatter}
\bmsection{Funding}
Carl Zeiss Foundation, Carl-Zeiss-Stiftung (CZS) via CZS Center for Quantum Photonics QPhoton.

\bmsection{Disclosures}
The authors declare no conflicts of interest.

\bmsection{Data Availability Statement}
No experimental data were generated or analyzed in this study. Derivations supporting the findings are provided in the article. Mathematica notebooks used for symbolic and numerical checks are available from the authors upon reasonable request.

\end{backmatter}

\bibliography{sample}

\end{document}